\documentclass[final            
  ]
  {aipproc}

\layoutstyle{6x9}

\newcommand{\ra}{\rightarrow}
\newcommand{\vub}{|V_{ub}|}
\newcommand{\vcb}{|V_{cb}|}
\newcommand{\btodslnu}{\bar{B}\ra D^* \ell\bar{\nu}}
\newcommand{\btopilnu}{\bar{B}\ra \pi\ell\bar{\nu}}
\newcommand{\lone}{\lambda_1}
\newcommand{\ltwo}{\lambda_2}
\newcommand{\Lbar}{\bar{\Lambda}}
\newcommand{\btoclnu}{b\ra c\ell\bar{\nu}}
\newcommand{\btoulnu}{b\ra u\ell\bar{\nu}}
\newcommand{\btosgamma}{b\ra s\gamma}

\begin{document}
\vbox{\hfill{CLNS 03/1844}}
\title{CLEO results on $|V_{cb}|$ and $|V_{ub}|$
\footnote{Presented at the Conference on the Intersections of
Particle and Nuclear Physics, CIPANP 2003.}
}

\author{Karl M.~Ecklund}{
  address={Floyd R.~Newman Laboratory for Elementary-Particle Physics\\
Cornell University, Ithaca, New York, 14853}
}

\begin{abstract}
I report results from the CLEO collaboration on semileptonic $B$
decays, highlighting measurements of the Cabibbo-Kobayashi-Maskawa
matrix elements $\vcb$ and $\vub$. I describe the techniques used to
obtain the recent improvements in precision for these measurements,
including the use of the $b\ra s\gamma$ photon spectrum to reduce
hadronic uncertainties in semileptonic $B$ decays.
\end{abstract}

\maketitle

The study of semileptonic $B$ meson decays allows measurement of the
Cabibbo-Kobayashi-Maskawa (CKM) matrix elements $\vcb$ and $\vub$,
providing important inputs to a test of the unitarity of the CKM
matrix, which governs the weak charged current and gives rise to $CP$
violation in the standard model.
The rate for a $b$ hadron to decay weakly to hadrons containing a $c$
or $u$ quark is proportional to $\vcb^2$ or $\vub^2$ respectively.
The absence of final-state interactions in semileptonic decay make the
interpretation less dependent on hadronic matrix elements than fully
hadronic $B$ decays, although hadronic uncertainties still limit the
precision of $\vub$ and $\vcb$ measurements.

The current round of measurements from CLEO continues to test the
hadronic calculations needed to disentangle weak matrix elements
from strong interaction effects.
For decays of $B$ mesons to exclusive final states, the hadronic
effects are expressed in terms of a form factor that depends only on
the momentum transfer $q^2$ to the lepton neutrino pair.  By measuring
decay rates as a function of $q^2$ we have begun to test the form
factors, particularly for $\btoulnu$ transitions.
In decays to inclusive final states, under the assumption of
parton-hadron duality, quark-level calculations may be compared to
inclusive measurements to extract CKM matrix elements.  Measurement of
spectral distributions in inclusive decays gives additional
observables to overconstrain theory parameters and test how well the
theory and parton-hadron duality works.

\subsection{$\vcb$ Measurements}
CLEO has measured $\vcb$ using the decay $\btodslnu$
\cite{Adam:2002uw}, where the decay rate as a function of $q^2$ is
extrapolated to maximum $q^2$ where the $D^*$ is at rest in the frame
of the initial $B$ meson.  At this kinematic point the form factor
${\cal F}$ is known to 4\% of itself, owing to heavy quark
symmetry considerations \cite{Falk:1993wt}.
The differential decay rate is given by
${d\Gamma\over dq^2}  = { G_F^2 \over 48 \pi^3} {|V_{cb}|^2}
                     { \left[{\cal F}(q^2)\right]^2} {{\cal K}(q^2)}$,
where $\cal K$ is a known kinematic function.
Using ${\cal F}(q^2_{\rm max})=0.91\pm0.04$ \cite{Hashimoto:2001nb}, 
we find $\vcb=(47.4 \pm 1.4_{\rm stat} \pm 2.0_{\rm syst} \pm 2.1_{\cal F})
\times 10^{-3}$,
somewhat higher than other results from $\btodslnu$.
The present world average from $\btodslnu$ is 
$\vcb=(42.4\pm 1.2_{\rm expt} \pm 1.9_{\rm theo})\times 10^{-3}$ 
\cite{HFAG:2003}.

A measurement of $\vcb$ using the inclusive semileptonic decay rate is
also possible.  Here the experimental inputs are the branching
fraction for $\bar{B}\to X_c\ell\bar{\nu}$ and the $B$ lifetime.  The
inclusive decay rate $\Gamma_c^{SL} = \gamma_c \vcb^2$, where
$\gamma_c$ comes from theory.
Within the framework of heavy quark effective theory (HQET)
\cite{ManoharWise:2000dt}, the inclusive semileptonic
decay rate is expanded in a double series in $\alpha_s^n$ and
$1/M^n$, where $M$ is the heavy quark mass.  Hadronic effects enter
both in the perturbative expansion and as expansion parameters, matrix
elements of non-perturbative QCD operators.  At ${\cal O}(1/M^2)$
there are two parameters: $\lone$, which is proportional to the
kinetic energy of the $b$ quark in the $B$ meson, and $\ltwo$,
which comes from the chromomagnetic operator.  An additional parameter
$\Lbar$ relates the $B$ meson mass to the $b$ quark mass.
From the $B$-$B^*$ mass difference $\ltwo=0.128\pm0.010$ GeV$^2$.
The other parameters can be estimated (\textit{e.g.}~in quark models) but
they can also be measured using spectral moments in inclusive $B$ decay.
Moments, \textit{e.g.}~of the lepton energy spectrum, are also
computed in HQET, allowing extraction of $\lone$ and $\Lbar$.

\begin{figure}
\includegraphics[height=5cm]{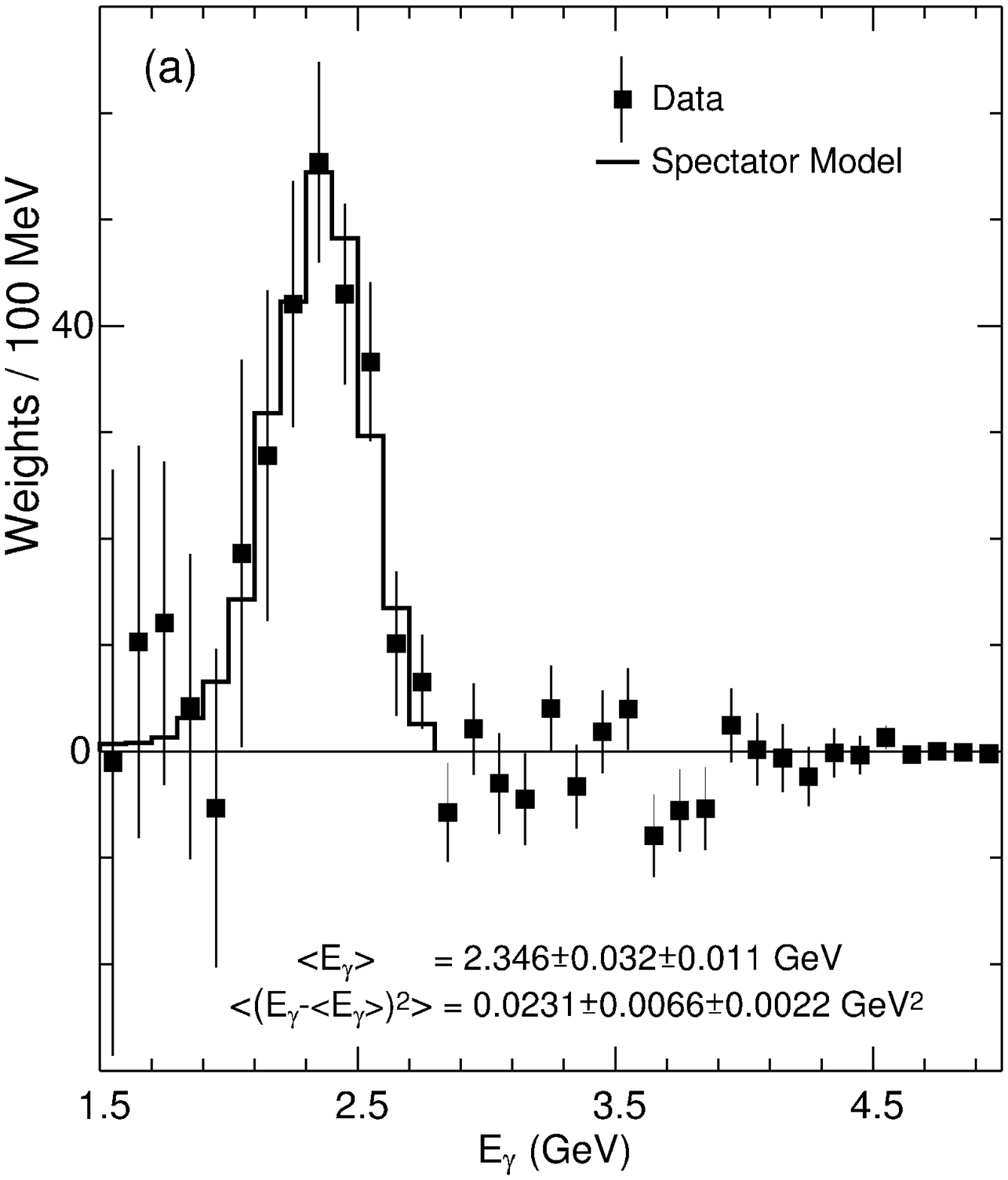}
\includegraphics[height=5cm]{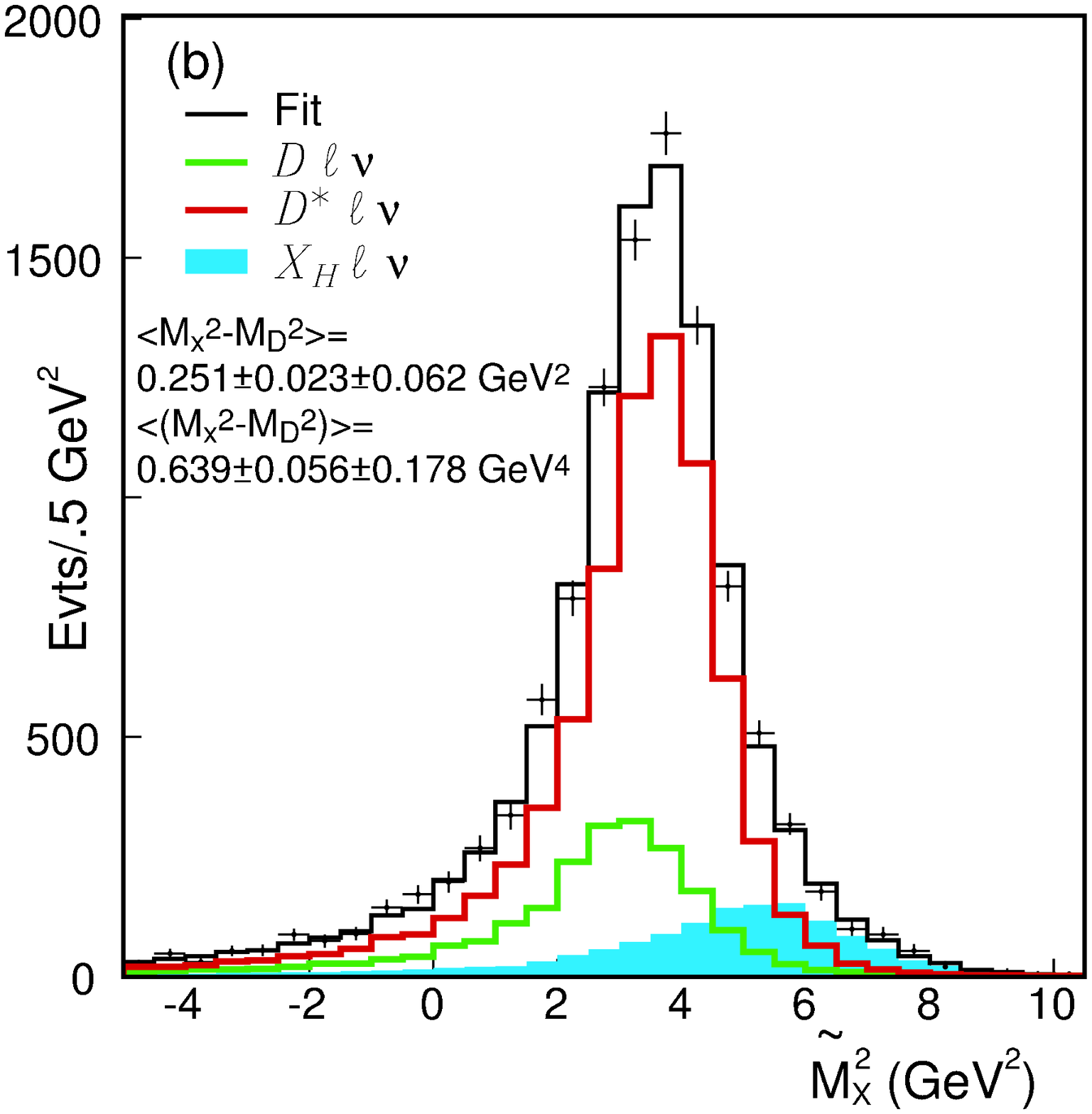}
\includegraphics[height=5cm,clip=true]{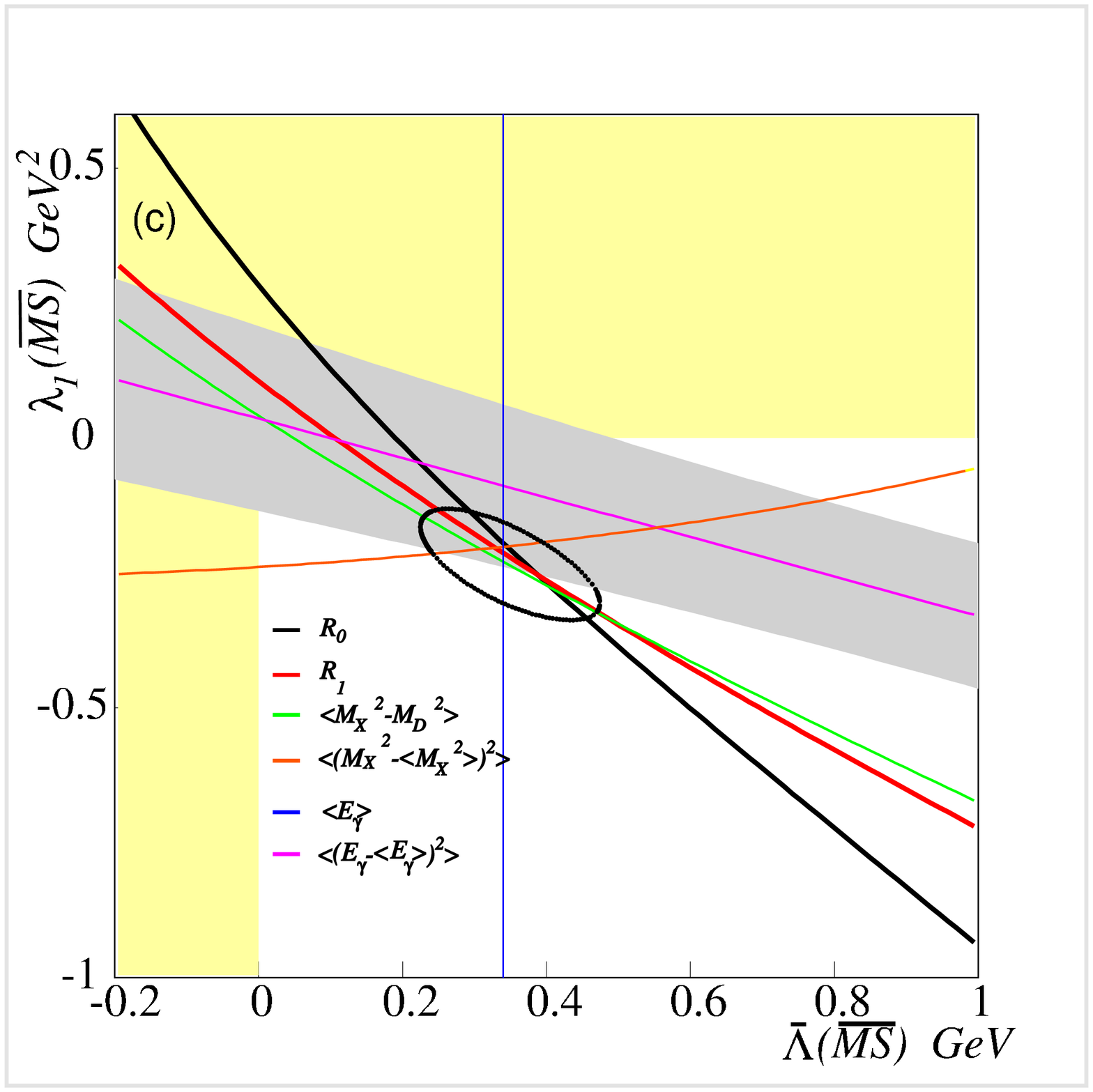}
\caption{$B\to X_s\gamma$ photon spectrum (a), $\bar{B}\to X_c\ell\bar{\nu}$ 
$M_X^2$ spectrum (b), and constraints on HQET parameters (c) from CLEO moment
measurements.  The shaded band includes ${\cal O}(1/M^3)$ theory
uncertainties.}
\label{kme:fig.hqet}
\end{figure}

CLEO has a preliminary measurement of the inclusive semileptonic branching
fraction using a high-momentum ($p>1.5$ GeV/$c$) lepton tag.  
The analysis is an update of Ref.~\cite{Barish:1996cx}.  The tag 
identifies a sample of $B$ decays with high purity (98\%).  Additional
electrons may come from the decay chain of the same $B$ or from the
decay of the other $B$ meson in the event 
($e^+e^-\ra\Upsilon(4S)\ra B\bar{B}$). 
Secondary leptons ($b\to c\to \ell$) and primary leptons are separated
using kinematic and charge correlations, with a known correction from
$B^0$-$\bar{B^0}$ mixing.  The new semileptonic branching fraction is 
$10.88\pm0.08\pm0.33$\%.  The spectrum of electrons above 600 MeV
is also obtained, from which spectral moments will be measured.

CLEO has recently measured spectral moments in inclusive semileptonic
decay and in $B\to X_s\gamma$.  These are used to extract HQET
parameters and reduce the theoretical uncertainty in inclusive $\vcb$
measurements.
CLEO measured the $B\to X_s\gamma$ photon spectrum and moments
(Fig.~\ref{kme:fig.hqet}a) in \cite{Chen:2001fj}.  
In \cite{Cronin-Hennessy:2001fk}, CLEO measured the moments of the
hadronic mass distribution in $\bar{B}\to X_c\ell\bar{\nu}$ decays
(Fig.~\ref{kme:fig.hqet}b). 
Combining the constraints on $\lone$ and $\Lbar$ from the first
moments of the photon energy and hadronic mass spectra, we obtain a solution
for $\lone$ and $\Lbar$ and extract 
$\vcb = (41.1 \pm 0.5_{\lone,\Lbar} 
              \pm 0.7_{\Gamma}
              \pm 0.8_{HQET}) \times 10^{-3}$
using the new CLEO branching fraction and PDG2003 lifetime average
as inputs.  The uncertainties from unknown ${\cal O}(1/M^3)$ HQET
parameters are dominant.

The lepton energy moments in $\bar{B}\to X_c\ell\bar{\nu}$ are also
sensitive to the HQET parameters, and CLEO has measured the lepton
spectrum \cite{Mahmood:2002tt} and moments \cite{Gremm:1996yn} above
1.5 GeV. From all of the moment measurements, one can assemble the
constraints on the HQET parameters $\lone$ and $\Lbar$.
Figure~\ref{kme:fig.hqet}c shows the remarkable consistency of these
measurements, lending credibility to the inclusive $\vcb$ measurement.

At present the inclusive $\vcb$ measurement is more precise (3\%) than
that from $\btodslnu$, but with reliance on HQET for hadronic
corrections.  The first tests of HQET using spectral moments in
inclusive $B$ decays give us some confidence in the method, but
additional tests with more inclusive moments are needed.  

The agreement between inclusive and exclusive measurements is another
test of our control of hadronic corrections.  There is good agreement
between inclusive and the world average exclusive $\vcb$ measurements,
but CLEO's exclusive $\vcb$ is larger than the inclusive measurement
and other measurements using $\btodslnu$.

\subsection{$\vub$ Measurements}
Measurements of $\btoulnu$ have to contend with a 50--100 times larger
background from $\btoclnu$.  Requiring a lepton energy above
the endpoint for $\btoclnu$ ($\approx 2.3$ GeV) is the easiest
strategy to reduce background, but this cut near the edge of the
spectrum introduces sensitivity to the motion of $b$ quark in the $B$
meson.  The sensitivity is reduced by using the $\btosgamma$ photon
spectrum \cite{Chen:2001fj}, which is sensitive to the same hadronic
effects at leading order 
\cite{Neubert:1994ch,Neubert:1994um,Bigi:1994ex,Leibovich:1999xf}.

CLEO measured the lepton spectrum from $B$ decays in the endpoint
region $E>2.2$ GeV and extracted a partial branching fraction of
$(2.30 \pm 0.15 \pm 0.35) \times 10^{-4}$ \cite{Bornheim:2002du}.
From the $\btosgamma$ photon spectrum, the fraction of $\btoulnu$
events passing the lepton energy cut is $f_u = 0.130 \pm 0.024 \pm 0.015$.
This gives 
$|V_{ub}| = (4.08 \pm 0.34_{\rm exp} \pm 0.44_{f_u} \pm 0.16_{\Gamma}
\pm 0.24_{NLO}) \times 10^{-3}$, where the theoretical
uncertainties are $\Gamma$, from \cite{Hoang:1998hm,Uraltsev:1999rr},
and $NLO$, from sub-leading terms relating hadronic effects in
$\btoulnu$ and $\btosgamma$.


\begin{figure}
\includegraphics[height=5cm]{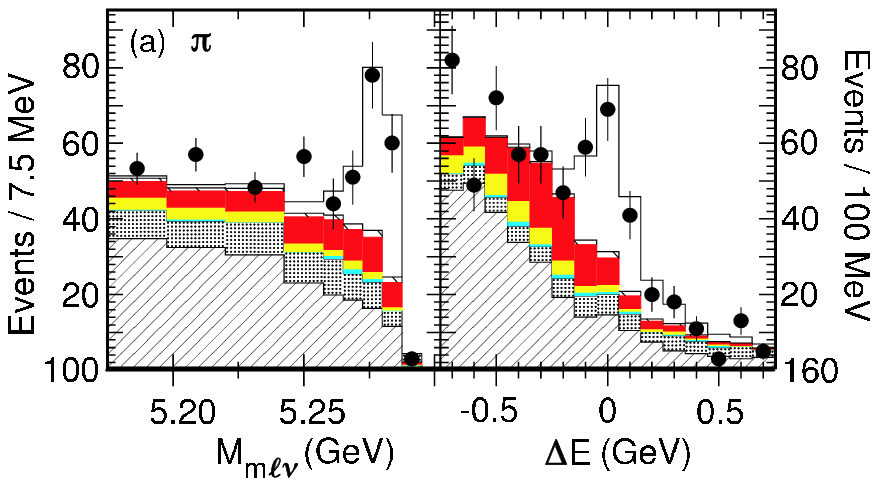}
\includegraphics[height=5cm]{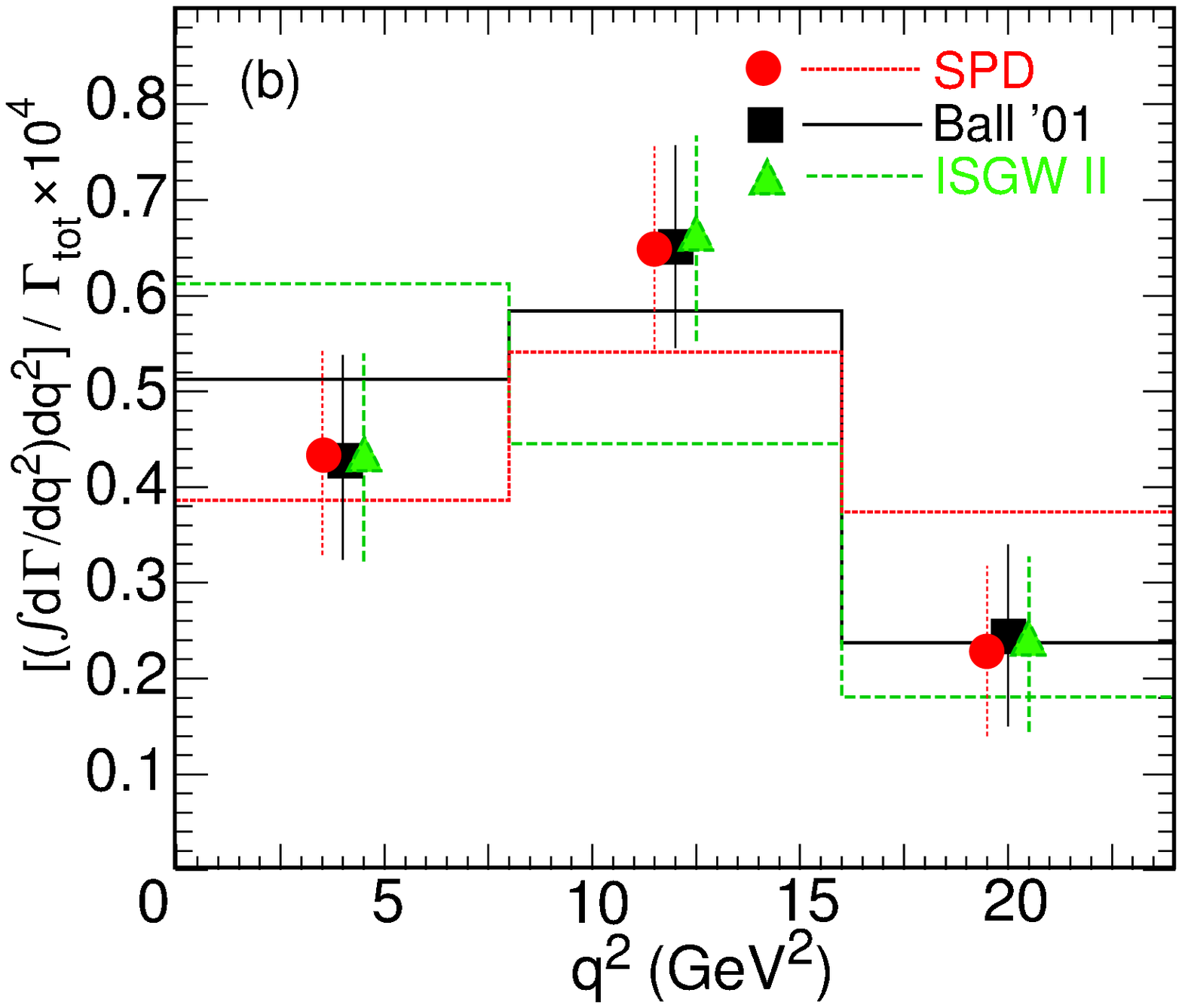}
\caption{Exclusive $\btopilnu$: (a) projections of ML fit to
$M_{m\ell\nu}$ and $\Delta E$ and (b) fit to $d\Gamma/dq^2$.}
\label{kme:fig.ExcVub}
\end{figure}

CLEO has also measured $\vub$ in the exclusive modes ${\bar B}\to
[\pi/\rho/\omega/\eta]\ell\bar{\nu}$ \cite{Athar:2003yg}, 
where kinematics from full 
reconstruction of the final state gives the needed suppression of
$\btoclnu$.  The neutrino is reconstructed from the missing energy and
momentum of the event, taking advantage of CLEO's large solid angle
(95\%).  Combined with a lepton and light meson candidate, energy and
momentum conservation leads to signal peaks in $\Delta E = E - E_{\rm
beam}$ and $M_{m\ell\nu}$, with $S/B \approx 1$.  We perform a
simultaneous maximum likelihood fit in $\Delta E$ and $M_{m\ell\nu}$ to
seven sub-modes.  Signals for $\pi$ (Fig.~\ref{kme:fig.ExcVub}a)
and $\rho$ are extracted
separately in three $q^2$ bins.  Given form factors from theory, we
extract $\vub$ from a fit to $d\Gamma/dq^2$ (Fig.~\ref{kme:fig.ExcVub}b). 
Combining $\btopilnu$ and
$\bar{B}\to\rho\ell\bar{\nu}$ results we find 
{$|V_{ub}| =  (3.17 \pm 0.17           |_{\rm stat}
                    \ ^{+0.16}_{-0.17} |_{\rm syst}
                    \ ^{+0.53}_{-0.39} |_{\rm theo}
                    \pm 0.03           |_{\rm FF}   ) \times 10^{-3}$}.
This result uses form factors from Lattice QCD ($q^2 > 16$ GeV$^2$) and
light cone sum rules ($q^2 > 16$ GeV$^2$) where each are most reliable.
In a test of $\btopilnu$ form factors, ISGW2
\cite{Scora:1995ty} is disfavored (Fig.~\ref{kme:fig.ExcVub}b).

We find good agreement between measurements of $\vub$ using inclusive
and exclusive techniques.  The theoretical uncertainty on the form
factor normalization currently limits the precision of the exclusive
$\vub$ measurement.  In the future, unquenched Lattice QCD
calculations can improve the $\btopilnu$ form factor in a limited
region of $q^2$. The inclusive $\btoulnu$ measurement can be 
further improved with increased $\btosgamma$ statistics and better
phenomenological understanding of non-perturbative shape functions
for the $B$ meson \cite{Leibovich:2002ys,Bauer:2002yu,Neubert:2002yx}.  
Comparison between inclusive measurements that use
different kinematic cuts (more inclusive and away from the endpoint
region) will increase our confidence in inclusive $\vub$
measurements.  Since the principal background comes from $\btoclnu$,
better knowledge of the dominant semileptonic $B$ decays will improve
systematic errors for both inclusive and exclusive measurements.

\bibliographystyle{aipproc}   

\bibliography{refs}

\begin{thebibliography}{22}
\expandafter\ifx\csname natexlab\endcsname\relax\def\natexlab#1{#1}\fi
\providecommand{\enquote}[1]{``#1''}
\expandafter\ifx\csname url\endcsname\relax
  \def\url#1{\texttt{#1}}\fi
\expandafter\ifx\csname urlprefix\endcsname\relax\def\urlprefix{URL }\fi

\bibitem[Adam et~al.(2003)]{Adam:2002uw}
Adam, N.~E., et~al., \emph{Phys. Rev.}, \textbf{D67}, 032001 (2003).

\bibitem[Falk and Neubert(1993)]{Falk:1993wt}
Falk, A.~F., and Neubert, M., \emph{Phys. Rev.}, \textbf{D47}, 2965--2981
  (1993).

\bibitem[Hashimoto et~al.(2002)]{Hashimoto:2001nb}
Hashimoto, S., et~al., \emph{Phys. Rev.}, \textbf{D66}, 014503 (2002).

\bibitem[{Heavy Flavor Averaging Group}(2003)]{HFAG:2003}
{Heavy Flavor Averaging Group}, Updates of semileptonic results (2003), for
  summer conferences.

\bibitem[Manohar and Wise(2000)]{ManoharWise:2000dt}
Manohar, A.~V., and Wise, M.~B., \emph{Heavy quark physics}, Cambridge
  University Press, 2000, chap.~4.

\bibitem[Barish et~al.(1996)]{Barish:1996cx}
Barish, B., et~al., \emph{Phys. Rev. Lett.}, \textbf{76}, 1570--1574 (1996).

\bibitem[Chen et~al.(2001)]{Chen:2001fj}
Chen, S., et~al., \emph{Phys. Rev. Lett.}, \textbf{87}, 251807 (2001).

\bibitem[Cronin-Hennessy et~al.(2001)]{Cronin-Hennessy:2001fk}
Cronin-Hennessy, D., et~al., \emph{Phys. Rev. Lett.}, \textbf{87}, 251808
  (2001).

\bibitem[Mahmood et~al.(2003)]{Mahmood:2002tt}
Mahmood, A.~H., et~al., \emph{Phys. Rev.}, \textbf{D67}, 072001 (2003).

\bibitem[Gremm et~al.(1996)]{Gremm:1996yn}
Gremm, M., Kapustin, A., Ligeti, Z., and Wise, M.~B., \emph{Phys. Rev. Lett.},
  \textbf{77}, 20--23 (1996).

\bibitem[Neubert(1994{\natexlab{a}})]{Neubert:1994ch}
Neubert, M., \emph{Phys. Rev.}, \textbf{D49}, 3392--3398 (1994{\natexlab{a}}).

\bibitem[Neubert(1994{\natexlab{b}})]{Neubert:1994um}
Neubert, M., \emph{Phys. Rev.}, \textbf{D49}, 4623--4633 (1994{\natexlab{b}}).

\bibitem[Bigi et~al.(1994)]{Bigi:1994ex}
Bigi, I. I.~Y., et~al., \emph{Int. J. Mod. Phys.}, \textbf{A9}, 2467--2504
  (1994).

\bibitem[Leibovich et~al.(2000)]{Leibovich:1999xf}
Leibovich, A.~K., Low, I., and Rothstein, I.~Z., \emph{Phys. Rev.},
  \textbf{D61}, 053006 (2000).

\bibitem[Bornheim et~al.(2002)]{Bornheim:2002du}
Bornheim, A., et~al., \emph{Phys. Rev. Lett.}, \textbf{88}, 231803 (2002).

\bibitem[Hoang et~al.(1999)]{Hoang:1998hm}
Hoang, A.~H., Ligeti, Z., and Manohar, A.~V., \emph{Phys. Rev.}, \textbf{D59},
  074017 (1999).

\bibitem[Uraltsev(1999)]{Uraltsev:1999rr}
Uraltsev, N., \emph{Int. J. Mod. Phys.}, \textbf{A14}, 4641--4652 (1999).

\bibitem[Athar et~al.(2003)]{Athar:2003yg}
Athar, S.~B., et~al., \emph{Phys. Rev.}, \textbf{D} (2003), hep-ex/0304019, to
  appear.

\bibitem[Scora and Isgur(1995)]{Scora:1995ty}
Scora, D., and Isgur, N., \emph{Phys. Rev.}, \textbf{D52}, 2783--2812 (1995).

\bibitem[Leibovich et~al.(2002)]{Leibovich:2002ys}
Leibovich, A.~K., Ligeti, Z., and Wise, M.~B., \emph{Phys. Lett.},
  \textbf{B539}, 242--248 (2002).

\bibitem[Bauer et~al.(2002)]{Bauer:2002yu}
Bauer, C.~W., Luke, M., and Mannel, T., \emph{Phys. Lett.}, \textbf{B543},
  261--268 (2002).

\bibitem[Neubert(2002)]{Neubert:2002yx}
Neubert, M., \emph{Phys. Lett.}, \textbf{B543}, 269--275 (2002).

\end{thebibliography}

\end{document}